\documentclass{article}
\usepackage{subcaption}
\usepackage{spconf,amsmath,graphicx}

\title{JND-based Perceptual Optimization for Learned Image Compression}
\name{Feng Ding$^{\star}$ \qquad Jian Jin$^{ \dagger}$ \qquad Lili Meng$^{\star}$\qquad Weisi Lin$^{\dagger}$
\thanks{© 20XX IEEE. Personal use of this material is permitted. Permission from IEEE must be obtained for all other uses, in any current or future media, including reprinting/republishing this material for advertising or promotional purposes, creating new collective works, for resale or redistribution to servers or lists, or reuse of any copyrighted component of this work in other works.}
}
\address{$^{\star}$ School of Information Science and Engineering, \\Shandong Normal University, Jinan, 250014, China.\\$^{\dagger}$ School of Computer Science and Engineering, \\Nanyang Technological University, 639798, Singapore}
\begin{document}
\topmargin=0mm
%\ninept
%
\maketitle
\begin{abstract}

Recently, learned image compression schemes have achieved remarkable improvements in image fidelity (\emph{e.g.}, PSNR and MS-SSIM) compared to conventional hybrid image coding ones due to their high-efficiency non-linear transform, end-to-end optimization frameworks, \emph{etc}. However, few of them take the Just Noticeable Difference (JND) characteristic of the Human Visual System (HVS) into account and optimize learned image compression towards perceptual quality. To address this issue, a JND-based perceptual quality loss is proposed. Considering that the amounts of distortion in the compressed image at different training epochs under different Quantization Parameters (QPs) are different, we develop a distortion-aware adjustor. After combining them together, we can better assign the distortion in the compressed image with the guidance of JND to preserve the high perceptual quality. All these designs enable the proposed method to be flexibly applied to various learned image compression schemes with high scalability and plug-and-play advantages. Experimental results on the Kodak dataset demonstrate that the proposed method has led to better perceptual quality than the baseline model under the same bit rate.
% \par\
% {\bf\emph{Keywords—JND, Perceptual Image Compression, Deep Codec.}\rm}
\end{abstract}
\section{Introduction}
\label{Intro}
Huge volumes of images are being captured, transmitted, and stored due to the popularization of smart devices in our daily life. Image compression is a key technology to address huge-images-caused network jams and storage overflow. Traditional hybrid image compression techniques, like JPEG \cite{wallace1991jpeg}, BPG \cite{bpg}, VVC \cite{vvc} \emph{etc.}, are mainly composed with linear transform (\emph{e.g.}, DCT \cite{dct}, WT \cite{wt}), quantization, entropy coding \emph{etc.}, where the spatial redundancy of image are mainly removed by linear transform. With the rising of deep learning, many learned image compression schemes \cite{balle2016end, balle2018variational,cheng2020learned} were proposed and outperformed the hybrid ones in terms of PSNR/MSE and MS-SSIM. However, as they optimized the compressed images toward statistically high fidelity without considering the perceptually high quality, this led to the low perceptual quality of their compressed images under low bit rates and made them unfriendly to human perception.
\begin{figure}[t]
    \begin{minipage}[h]{0.49\linewidth}
         \centering
         Baseline
     \end{minipage}
     \hfill
     \begin{minipage}[h]{0.49\linewidth}
     \centering
         Ours
     \end{minipage}
     \vspace{3px}

    \begin{subfigure}[h]{0.49\linewidth}
        \includegraphics[width=1\linewidth]{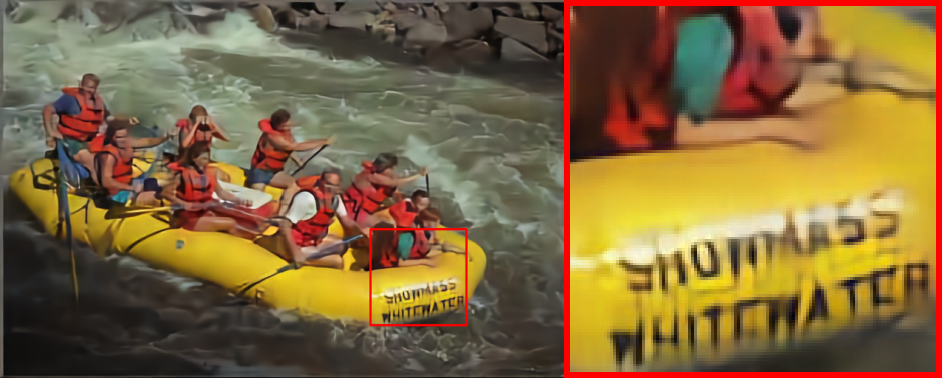}
        \centering
    \end{subfigure}
    \hspace{1px}
    \hfill 
    \begin{subfigure}[h]{0.49\linewidth}
        \includegraphics[width=1\linewidth]{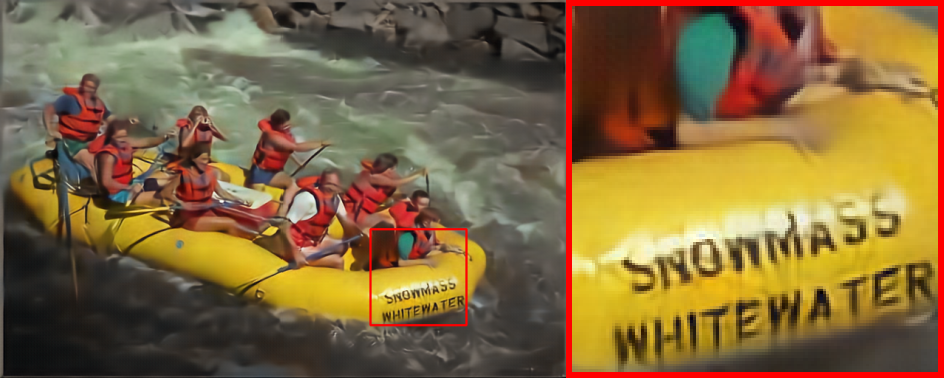}
        \centering
    \end{subfigure}
    \vspace{3px}
    
    \begin{subfigure}[h]{0.49\linewidth}
        \includegraphics[width=1\linewidth]{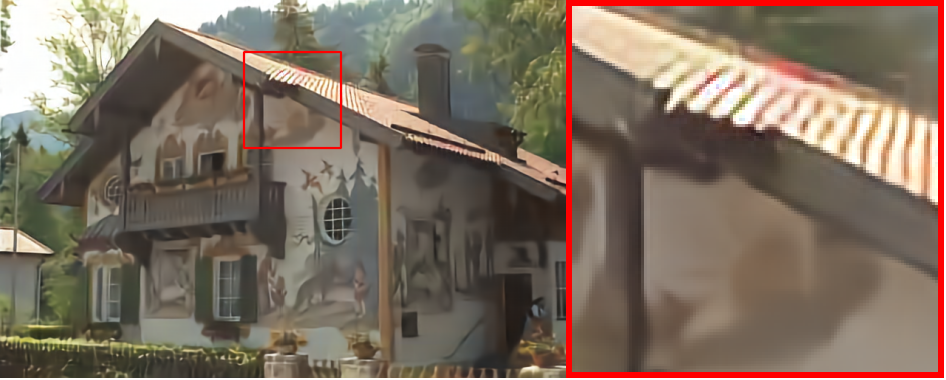}
        \centering
    \end{subfigure}
    \hfill 
    \begin{subfigure}[h]{0.49\linewidth}
        \includegraphics[width=1\linewidth]{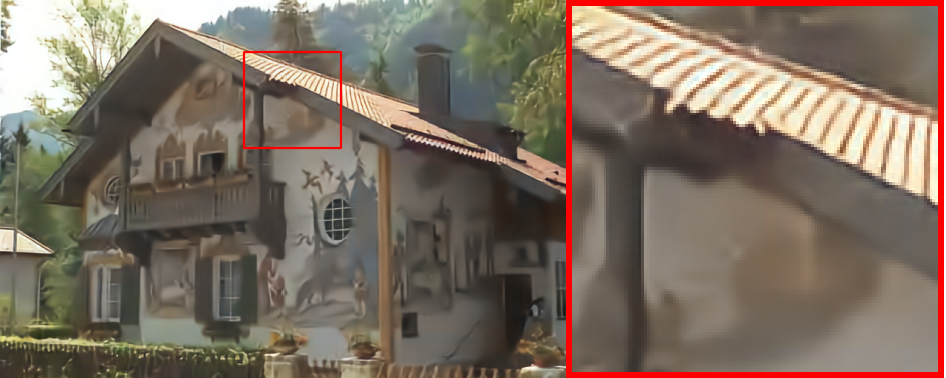}
        \centering
    \end{subfigure}
    \caption{The left column images are compressed with baseline \cite{balle2016end}, the right column images are compressed with our proposed method. The BPP of the top left, top right, bottom left, and bottom right images are 0.144, 0.151, 0.135, and 0.148, respectively.}\label{fig:0}
\end{figure}

% Perceptual image/video compression aims to exploit visual redundancies so as to maximize compression efficiency, which is widely used in the industry due to its high performance on perceptual quality. JND as one of the main characteristics of the HVS refers to the maximum visual changes that the HVS can't perceive, which is a metric to measure the visual redundancy of the image/video. Hence, JND was widely used in perceptual image/video compression. For instance, Yang \emph{et al.} \cite{yang2005just} proposed a pixel domain JND model to filter unperceived changes during motion estimation, which speeds up encoding and saves bits. After that, Bae \emph{et al.} \cite{bae2014novel} proposed a sub-band domain JND model, which was applicable to any size of transform kernel and its variant \cite{bae2016hevc} is further used in HEVC-based perceptually adaptive video coding. Then, a picture-wise JND model \cite{liu2019deep} was proposed and used for Quality Factor (QF) selection during image compression. Recently, several learning-based JND models \cite{wu2020unsupervised, jin2021just, jin2022full,jiang2022towards,jin2022hvs} further improved the accuracy of JND prediction, which provided new techniques of perceptual image/video compression. JND-based perceptual hybrid image/video compression has been fastly developed in recent years. How to design a feasible solution to achieve perceptual optimization for learned image compression based on JND is still an open problem.

Perceptual image/video compression aims to exploit visual redundancies so as to maximize compression efficiency, which is widely used in the industry due to its high performance on perceptual quality. JND as one of the main characteristics of the HVS refers to the maximum visual changes that the HVS can't perceive, which is a metric to measure the visual redundancy of the image/video. Hence, JND was widely used in traditional perceptual image/video compression. But how to design a feasible solution based on JND to achieve perceptual optimization for learned image compression is still an open problem. 
% %For instance, Yang \emph{et al.} \cite{yang2005just} proposed a pixel domain JND model to filter unperceived changes during motion estimation, which speeds up encoding and saves bits. After that, Bae \emph{et al.} \cite{bae2014novel} proposed a sub-band domain JND model, which was applicable to any size of transform kernel and its variant \cite{bae2016hevc} is further used in HEVC-based perceptually adaptive video coding. Then, a picture-wise JND model \cite{liu2019deep} was proposed and used for Quality Factor (QF) selection during image compression. 
% Recently, several learning-based JND models \cite{wu2020unsupervised, jin2021just, jin2022full,jiang2022towards,jin2022hvs} further improved the accuracy of JND prediction, which provided new techniques of perceptual image/video compression. Although JND-based perceptual hybrid image/video compression has been fastly developed in recent years, how to design a feasible solution based on JND to achieve perceptual optimization for learned image compression is still an open problem.
Recently, several learning-based JND models \cite{wu2020unsupervised, jin2021just, jin2022full,jiang2022towards,jin2022hvs} further improved the accuracy of JND prediction, which provided new techniques and perspectives of perceptual image/video compression. For instance, in \cite{jin2022full}, the authors have demonstrated that high perceptual results can be achieved even when 10 times JND (PSNR$=$26.06dB) is injected into the original image. This is because amounts of changes are assigned to the insensitivity regions/channels, achieving the minimum perceptual loss. In other words, times of JND can provide good guidance for distortion assignment to preserve the high perceptual quality of compressed images even under low bit rates.

In view of the observation above, we propose a JND-based perceptual optimization for learned image compression. To this end, we developed a JND-based perceptual quality loss together with a distortion-aware adjustor to cope with different amounts of distortion at different training epochs under different QPs. For the small amounts of distortion (below or equal to the JND), the adjustor would be set to 1. In this case, the JND will be used to induce the distortions being assigned to the insensitivity regions of the compressed images so that such distortion can not be perceived by the HVS. For large amounts of distortion, the adjustor would be determined by the amount of distortion. In this case, as the amount of distortion is largely beyond the JND, JND would be limited for guiding such large distortion distribution, while times of JND will be used to guide the distortions assignment so that such distortion will cause minimum perceptual loss for the HVS. All these designs above enable the proposed method to be flexibly applied to the learned image compression schemes and achieve better perceptual quality, as shown in Fig. \ref{fig:0}.

\begin{figure}[t]
\centering
\includegraphics[width=.45\linewidth]{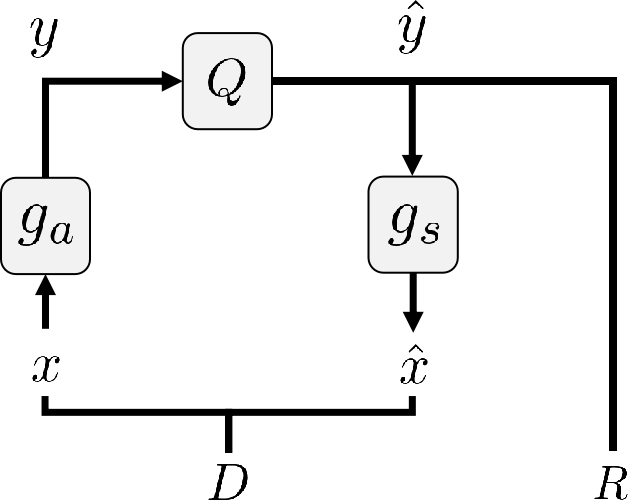}
\DeclareGraphicsExtensions.
\caption[]{Framework of the baseline method in \cite{balle2016end}. $x$ and $\hat{x}$ are source image and its associated decoded image.}
\label{fig:1}
\end{figure}
\begin{figure*}[ht!]
    \centering
    \begin{subfigure}[h]{0.3\linewidth}
        \includegraphics[width=1\linewidth]{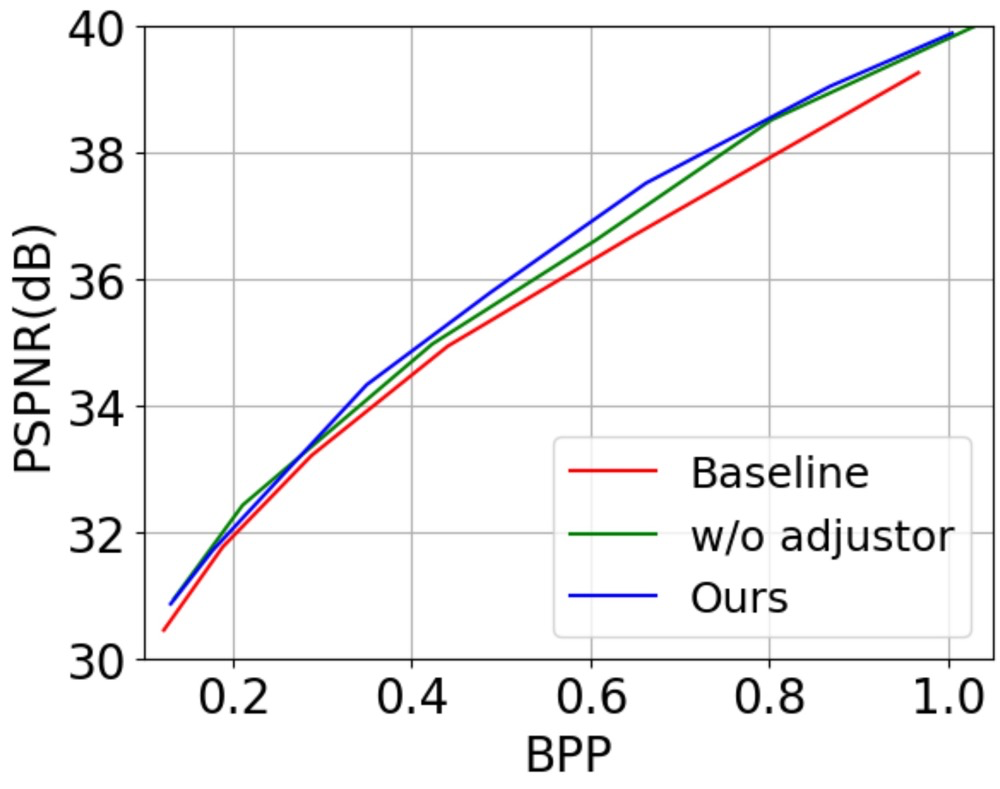}\caption{}\label{fig:3.1}
    \end{subfigure}
    \hfill
    \begin{subfigure}[h]{0.3\linewidth}
        \includegraphics[width=1\linewidth]{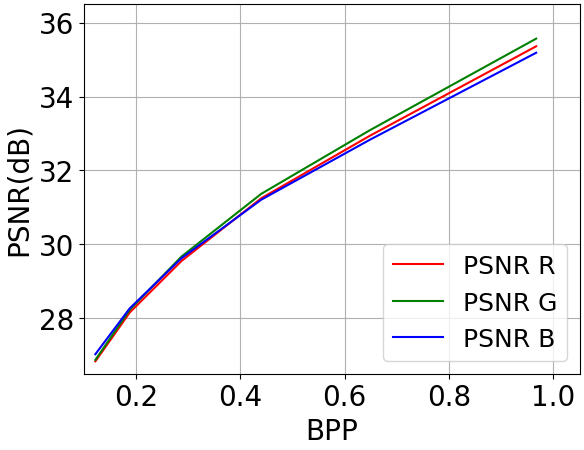}\caption{}\label{fig:3.2}
    \end{subfigure}
    \hfill
    \begin{subfigure}[h]{0.3\linewidth}
        \includegraphics[width=1\linewidth]{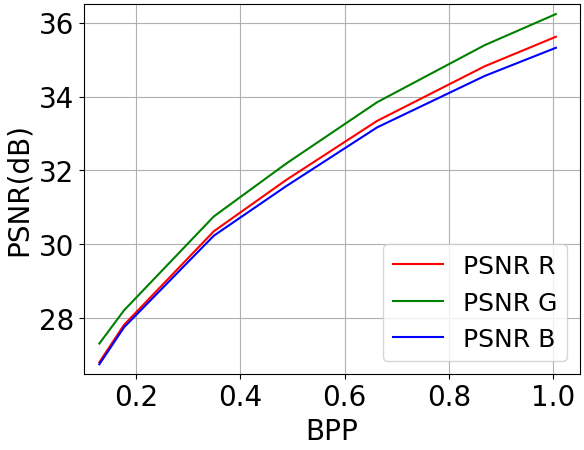}\caption{}\label{fig:3.3}
    \end{subfigure}
    \caption{(\subref{fig:3.1}) shows the PSPNR-BPP curves of the baseline and the proposed method. (\subref{fig:3.2}) and (\subref{fig:3.3}) show the PSNR-BPP curves of the baseline and the proposed method on the red, green, and blue channels.}\label{fig:3}
\end{figure*}
\section{Methods}\label{sec:2}
The proposed JND-based perceptual optimization can be applied to any learned image compression scheme. For simplicity, we start with a typical learned image compression framework in \cite{balle2016end} and briefly review it as our baseline. Then, a JND-based perceptual quality loss is proposed to replace the MSE of the baseline for perceptual optimization. Meanwhile, a distortion-aware adjustor is developed and incorporated with the JND-based perceptual quality loss. Finally, the detailed training processes are introduced.

\subsection{Framework of Learned Image Compression}
A typical framework of the learned image compression scheme is shown in Fig. \ref{fig:1}. Image $x$ is firstly represented with latent code via a parametric analysis transform $g_a$. After that, the latent code is quantized to a discrete-valued vector $\hat{y}$ via quantizer $Q$, which is then compressed as a stream and the rate cost $R$ is obtained. Then, the stream is decoded to $\hat{y}$. After it fed into a  parametric synthesis transform $g_s$, an image $\hat{x}$ is reconstructed. To make sure that the reconstructed image $\hat{x}$ has a high signal fidelity, MSE is used to optimize distortion between $x$ and $\hat{x}$. This whole process can be represented as
\begin{equation}
\begin{aligned}
\small
& y = g_a(x,\theta_a)\\
& \hat{y} = Q(y)\\
& \hat{x} = g_s(\hat{y},\theta_s),
\end{aligned}\label{eq1}
\end{equation}
where $\theta_a$ and $\theta_s$ are the optimized weights of analysis transform network $g_a$ and synthesis transform network $g_s$. Then, the objective function of the learned image compression is formulated as follows
\begin{equation}
\begin{aligned}
\small
L & = R(\hat{y})+\lambda D(x,\hat{x}), \text{where}\ D(x,\hat{x}) = \text{MSE}(x,\hat{x}).
\end{aligned}\label{eq2}
\end{equation}
$\lambda$ is a Lagrange multiplier to trade-off between the rate $R$ and distortion $D$. Distortion $D$ is evaluated by the MSE between original image $x$ and its associated reconstructed one $\hat{x}$. In other words, all the differences between $x$ and $\hat{x}$ are counted as the distortion in the learned image compression. 

\subsection{JND-based Perceptual Quality Loss}
According to the JND concept introduced in Sec. \ref{Intro}, only the distortion beyond the JND threshold is able to be perceived by the HVS and cause visual degradation. In the meantime, the distortion under the JND threshold isn't perceived by the HVS and doesn't affect the perceptual quality. In view of this, we define JND-based perceptual quality loss $D(x,\hat{x},j)$ as follows
\begin{equation}
\begin{aligned}
\small
D(x,\hat{x},j) = ReLU(d(x(h,w,c),\hat{x}(h,w,c))-\\\alpha \cdot j_o(h,w,c))^{2},
\end{aligned}\label{eq3}
\end{equation}
$\text{where}\ d(x(h,w,c),\hat{x}(h,w,c)) = |x(h,w,c)-\hat{x}(h,w,c)|$. $d$ is the absolute difference between two pixels with the same location $(h,w,c)$ in $x$ and $\hat{x}$. $j_o$ is the JND map generated with the method in \cite{jin2022full}. $\alpha$ is the distortion-aware adjustor, to be introduced in \ref{sec:2.3}. $ReLU$ is an activation function. %With such a loss function, more distortion can be assigned in the insensitivity regions of compressed images.  

By replacing the MSE loss used in Eq. \eqref{eq2} with the proposed JND-based perceptual quality loss, we can optimize the learned image compression toward perceptual quality.

\subsection{Distortion-aware Adjustor}\label{sec:2.3}
For a learned image compression scheme, we usually train several QPs to fit the different requirements of bandwidth, which leads to different amounts of distortion in the compressed images. Besides, even for the same QP at different epochs during the training process, the amount of distortion keeps changing. In view of this, we propose a distortion-aware adjustor, i.e., $\alpha$ in Eq. \eqref{eq3}. Here, we use $\alpha_{e,q}$ to represent the distortion-aware adjustor at epoch $e$ under QP $q$. Then, $\alpha_{e,q}$ is defined as follows 
\begin{equation}
\small
\begin{split}
\alpha_{e,q} =\left\{\begin{array}{l}
\frac{\sum_{h,w,c}{d(h,w,c)}}{\sum_{h,w,c}{j_o(h,w,c)}}, \text{if}\ \frac{\sum_{h,w,c}{d(h,w,c)}}{\sum_{h,w,c}{j_o(h,w,c)}} > 1, \\
1, \text{otherwise.}
\end{array}\right. 
\end{split}
\label{eq4}
\end{equation}
% The amount of distortion in a compressed image $\hat{x}_{e,q}$ at epoch $e$ under QP $q$ is denoted by $d(x(h,w,c),\hat{x}_{e,q}(h,w,c))$. If $\hat{x}_{e,q} > 1$, a large amount of distortion is present, and $\hat{x}_{e,q}$ times of JND will guide the distortion assignment to achieve minimum perceptual loss. If $\hat{x}_{e,q} \leq 1$, JND will guide the distortion assignment as the amount of distortion is below or equal to JND. All designs aim to maintain the perceptual quality of compressed images. The JND-based perceptual optimization objective function can be rewritten accordingly.

where $d(h,w,c)=d(x(h,w,c),\hat{x}_{e,q}(h,w,c))$ denotes the amount of distortion in compressed image $\hat{x}_{e,q}$ at epoch $e$ under QP $q$. If $\hat{x}_{e,q} > 1$, it means that a large amount of distortion exists in the compressed image. In this case, $\hat{x}_{e,q}$ times of JND will be used to guide the distortion assignment so that the compressed image achieves the minimum perceptual loss. If $\hat{x}_{e,q} \leq 1$, it means that the amount of distortion in the images below or equal to the JND, then JND will be used for guiding the distortion assignment. All these designs try to preserve the perceptual quality of the compressed images. Then, the objective function of JND-based perceptual optimization can be rewritten as
\begin{equation}
\begin{aligned}
L & = R(\hat{y})+\lambda D(x,\hat{x}_{e,q},\alpha_{e,q}, j_o).
\end{aligned}\label{eq5}
\end{equation}

\subsection{Training Strategy}\label{sec:2.4}
In this work, the well-trained model of the baseline is loaded as pre-trained parameters for training the proposed perceptual optimized one. Besides, we use a pre-trained model with low QP to train our high QP model, since the low QP model has more detailed information and can achieve perceptual optimization by only discarding the information that is insensitive to the HVS. If we train our model based on the pre-trained model with the same QP, it requires recovering sensitive information while discarding insensitive one, which is hard to train. This training strategy reduces the bit rate while maintaining the same perceptual quality, as demonstrated by ablation experiments in Sec. \ref{sec:3.2}.
\section{Experiments}\label{sec:3}
\textbf{Datasets.} CLIC \cite{toderici2020workshop} and DIV2K \cite{agustsson2017ntire} datasets are selected as the training set, where all the images in CLIC and DIV2K (including training set and validation set) are used for training. Each image is randomly cropped into 200 samples of size (256,256). We evaluate the performance of the baseline method and the proposed one on the Kodak \cite{kodak} dataset. 

\textbf{Settings.} We set the batch size and learning rate to 200 and 1e-4, and train our models under seven different QPs by setting $\lambda$ to 0.0033, 0.0063, 0.015, 0.025, 0.04, 0.065, and 0.085, respectively. The channel number of the bottleneck is set to 128 for the first four QPs and 192 for the rest of the three QPs, similar to the baseline settings. To reduce the learning rate when test loss has stopped improving, we use a plateau scheduler, where the learning rate is decreased by a factor of 10 if no improvement is seen for 2 epochs. The training process is completed when the learning rate drops to 1e-6.

\subsection{Experimental Results}\label{sec:3.1}
As there is no work utilizing JND to perceptually optimize the learned image compression, we compare the proposed method with the baseline \cite{balle2016end}. 

\textbf{Objective evaluation.} In the objective evaluation, the PSPNR \cite{bai2014multiple} metric (a variant of PSNR) that is widely used in traditional JND-based perceptual image and video coding is utilized here to evaluate the perceptual quality of the compressed images, which is calculated as $\text{PSPNR}=10\log_{10}\frac{255^2}{D(x,\hat{x},j)}$. $D(x,\hat{x},j)$ is obtained from Eq. \eqref{eq3}. Then, the PSPNR-BPP curves are shown in Fig. \ref{fig:3} (\subref{fig:3.1}), where the PSPNR-BPP curves of the baseline and the proposed method are in red and blue colors, respectively. A higher PSPNR means better perceptual quality. It can be obviously observed that the PSPNR-BPP curve of the proposed method is always above that of the baseline. That is, under the same BPP, the images compressed with the proposed method have better perceptual quality than the baseline. 

\begin{figure*}[ht!]
    \begin{minipage}[h]{0.24\linewidth}
         \centering
         Source Image
     \end{minipage}
     \hfill
     \begin{minipage}[h]{0.24\linewidth}
     \centering
         Baseline
         \end{minipage}
     \hfill
     \begin{minipage}[h]{0.24\linewidth}
     \centering
         Ours
     \end{minipage}
     \hfill
     \begin{minipage}[h]{0.24\linewidth}
     \centering
         Ablation study
     \end{minipage}
     \vspace{2px}
     
    \begin{subfigure}[h]{0.245\linewidth}
        \includegraphics[width=1\linewidth]{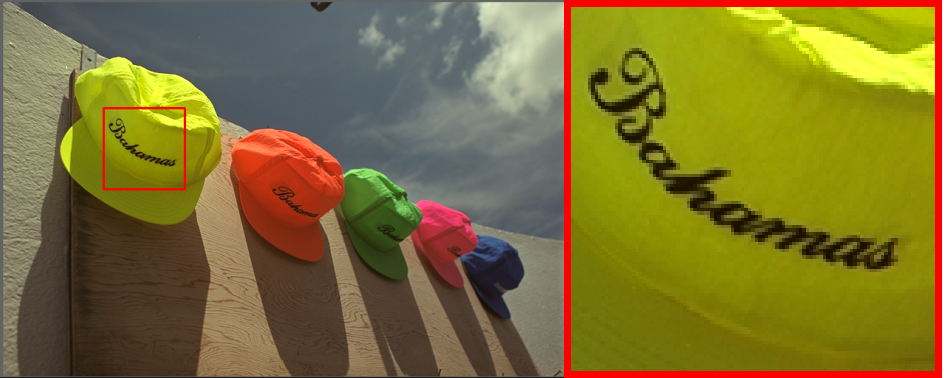}
        \centering    
        BPP\hspace{3px}/\hspace{3px}PSNR\hspace{3px}/\hspace{3px}MS-SSIM
    \end{subfigure}
    % \hspace{1px}
    \hfill
    \begin{subfigure}[h]{0.245\linewidth}
        \includegraphics[width=1\linewidth]{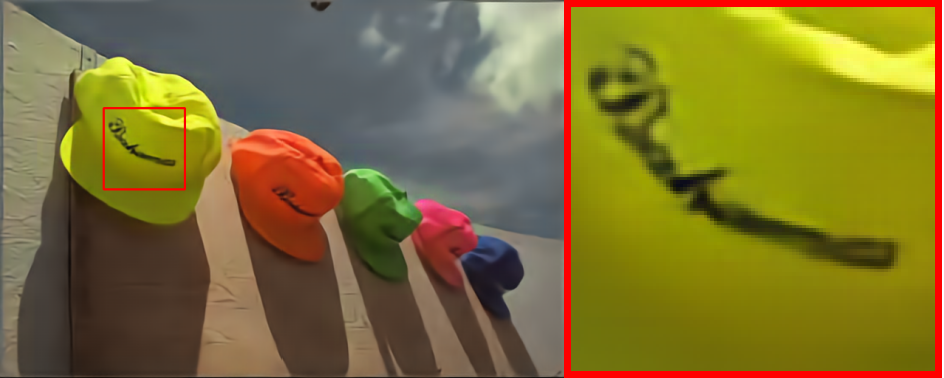}
        \centering
        0.091\hspace{3px}/\hspace{3px}30.141\hspace{3px}/\hspace{3px}0.940
    \end{subfigure}
    % \hspace{1px}
    \hfill
    \begin{subfigure}[h]{0.245\linewidth}
        \includegraphics[width=1\linewidth]{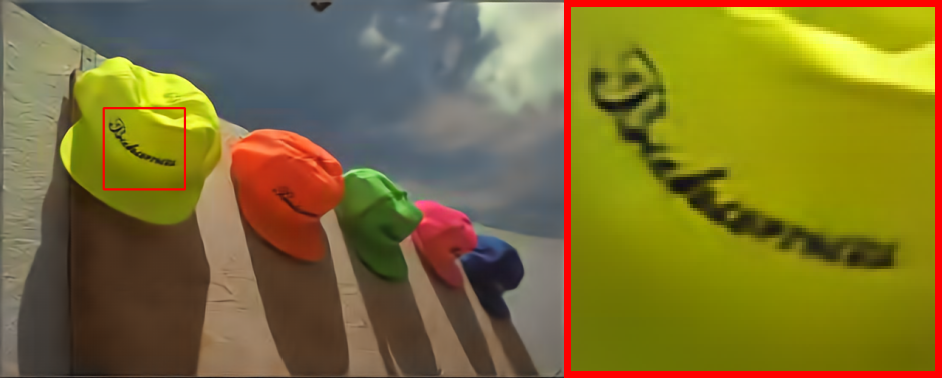}
        \centering
        0.091\hspace{3px}/\hspace{3px}29.695\hspace{3px}/\hspace{3px}0.933
    \end{subfigure}
    % \hspace{1px}
    \hfill
    \begin{subfigure}[h]{0.245\linewidth}
        \includegraphics[width=1\linewidth]{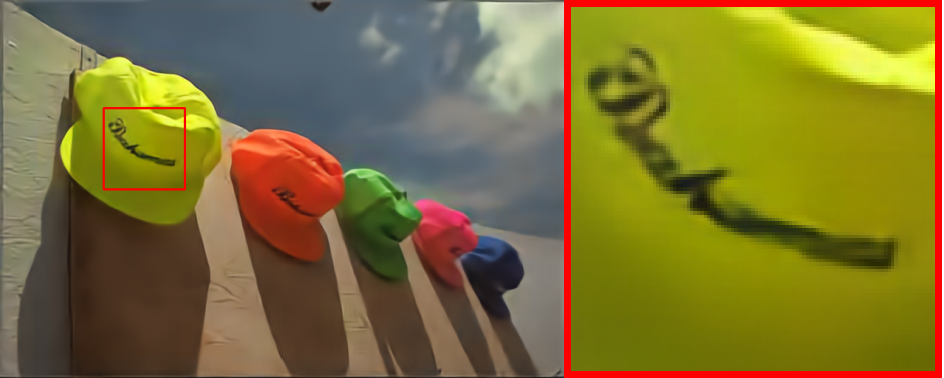}
        \centering
        0.086\hspace{3px}/\hspace{3px}29.584\hspace{3px}/\hspace{3px}0.935
    \end{subfigure}
    \vspace{3px}
    % \begin{subfigure}[h]{0.245\linewidth}
    %     \includegraphics[width=1\linewidth]{figures/Kodak14_o_v2.png}
    %     \centering
    %     BPP
        
    %     PSNR\hspace{5px}/\hspace{5px}MS-SSIM
    % \end{subfigure}
    % \hfill
    % \begin{subfigure}[h]{0.245\linewidth}
    %     \includegraphics[width=1\linewidth]{figures/Kodak14_b_v2.png}
    %     \centering
    %     0.144
        
    %     25.247\hspace{5px}/\hspace{5px}0.889
    % \end{subfigure}
    % \hfill
    % \begin{subfigure}[h]{0.245\linewidth}
    %     \includegraphics[width=1\linewidth]{figures/Kodak14_j_v2.png}
    %     \centering
    %     0.151
        
    %     25.163\hspace{5px}/\hspace{5px}0.861
    % \end{subfigure}
    % \hfill
    % \begin{subfigure}[h]{0.245\linewidth}
    %     \includegraphics[width=1\linewidth]{figures/Kodak14_a_v2.png}
    %     \centering
    %     0.138
        
    %     25.015\hspace{5px}/\hspace{5px}0.878
    % \end{subfigure}
    %     \vspace{5px}
    
    \begin{subfigure}[h]{0.245\linewidth}
        \includegraphics[width=1\linewidth]{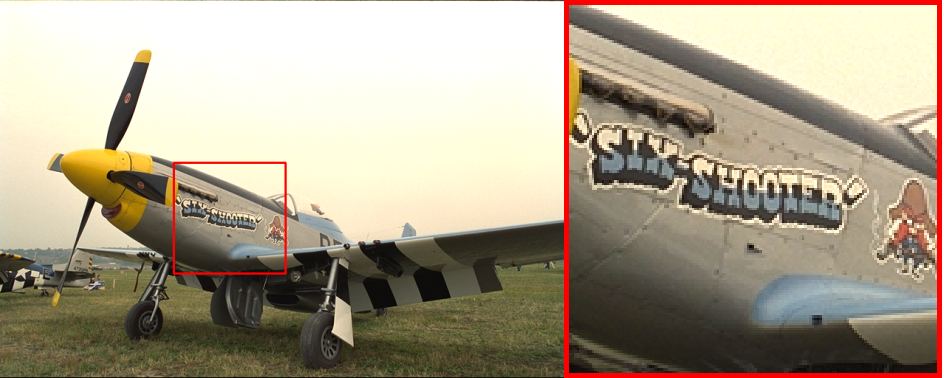}
        \centering
        BPP\hspace{3px}/\hspace{3px}PSNR\hspace{3px}/\hspace{3px}MS-SSIM
    \end{subfigure}
    \hfill
    \begin{subfigure}[h]{0.245\linewidth}
        \includegraphics[width=1\linewidth]{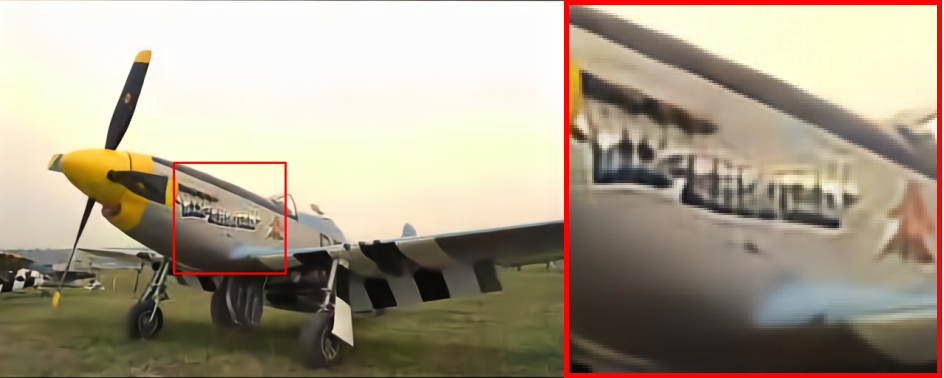}
        \centering
        0.100\hspace{3px}/\hspace{3px}28.869\hspace{3px}/\hspace{3px}0.945
    \end{subfigure}
    \hfill
    \begin{subfigure}[h]{0.245\linewidth}
        \includegraphics[width=1\linewidth]{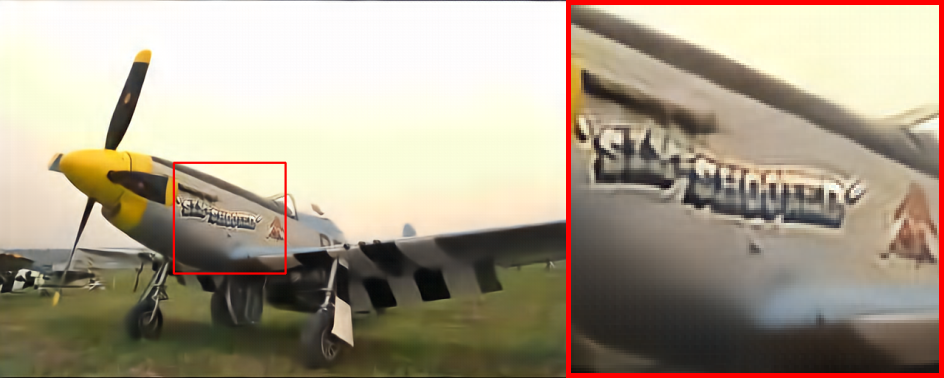}
        \centering
        0.108\hspace{3px}/\hspace{3px}28.823\hspace{3px}/\hspace{3px}0.936
    \end{subfigure}
    \hfill
    \begin{subfigure}[h]{0.245\linewidth}
        \includegraphics[width=1\linewidth]{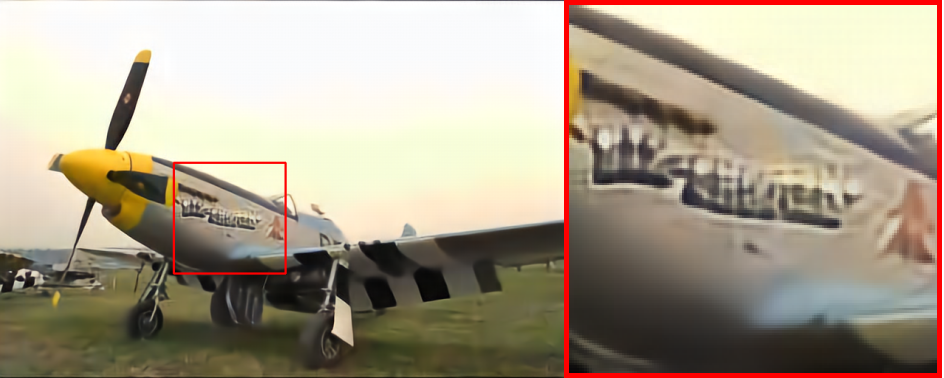}
        \centering
        0.096\hspace{3px}/\hspace{3px}28.544\hspace{3px}/\hspace{3px}0.940
    \end{subfigure}
    
    \caption{Images in the first column are the source images from the Kodak dataset. Images in the second and third columns are the images compressed by the baseline \cite{balle2016end} and our proposed method (trained with the proposed training strategy), respectively. The last column shows the image compressed by the proposed method (trained without the proposed training strategy).}\label{fig:2}
\end{figure*}

\textbf{Subjective evaluation.} To verify that the proposed method has better perceptual quality, we show the images compressed via the baseline and the proposed method with a similar bit rate in Fig. \ref{fig:2}. It can be seen that the perceptual quality of our method is significantly better than that of the baseline, especially in the numbers and text regions. Besides, we show the PSNR-BPP curves in three color channels in Fig. \ref{fig:3} (\subref{fig:3.2}) and (\subref{fig:3.3}). We can observe that the PSNR-BPP curves almost overlap in the results of the baseline, while they are discrete in our results. The PSNR of the green, red, and blue channels are high, medium, and low, respectively. This means our method assigns more distortion in the blue and red channels, which are insensitive to the HVS.

\subsection{Ablation Study}\label{sec:3.2}
\textbf{Ablation of distortion-aware JND adjustor.} To verify the performance of the proposed distortion-aware JND adjustor, we set the $\alpha$ in Eq. \eqref{eq5} to a constant value 10 here and retrain models, other settings are the same as our proposed method. The PSPNR-BPP curve is shown in Fig. \ref{fig:3} (\subref{fig:3.1}). It can be seen that the proposed method with the distortion-aware JND adjustor outperforms that without the distortion-aware JND adjustor in most of QPs.
\begin{figure}[t!]
    \begin{minipage}[h]{0.3\linewidth}
         \centering
         Source Image
     \end{minipage}
     \hfill
     \begin{minipage}[h]{0.3\linewidth}
         \centering
         Baseline Cheng
     \end{minipage}
     \hfill
     \begin{minipage}[h]{0.3\linewidth}
     \centering
         Ours
     \end{minipage}
     \vspace{3px}
    
    \begin{subfigure}[h]{0.3\linewidth}
        \includegraphics[width=1\linewidth]{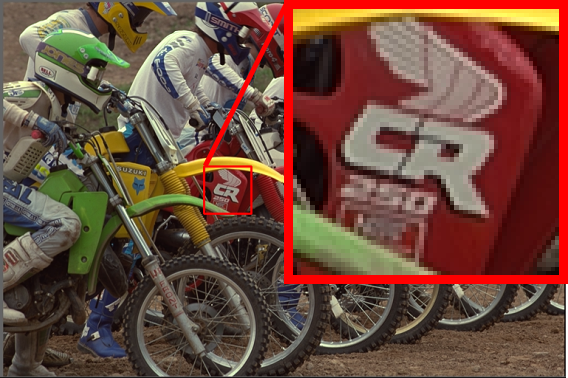}
        \centering
        BPP
        
        PSNR\hspace{2px}/\hspace{2px}MS-SSIM
    \end{subfigure}
    \hfill 
    \begin{subfigure}[h]{0.3\linewidth}
        \includegraphics[width=1\linewidth]{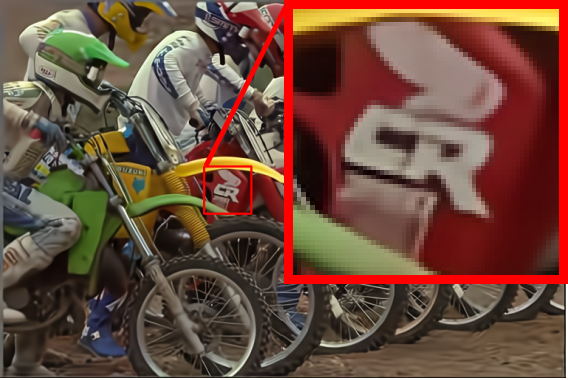}
        \centering
        0.236
        
        25.948\hspace{2px}/\hspace{2px}0.937
    \end{subfigure}
    \hfill
    \begin{subfigure}[h]{0.3\linewidth}
        \includegraphics[width=1\linewidth]{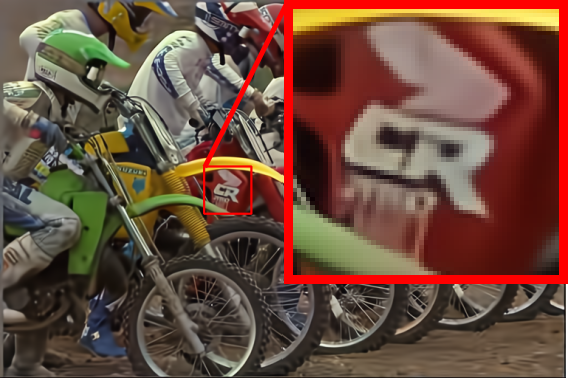}
        \centering
        0.233
        
        25.837\hspace{2px}/\hspace{2px}0.934
    \end{subfigure}
    \caption{The first image is a source image. The second and third images are the images compressed with Cheng \cite{cheng2020learned} and the proposed method, respectively. }\label{fig:4}
\end{figure}

\textbf{Ablation of the training strategy.} To verify the effectiveness of the training strategy in Sec. \ref{sec:2.4}, we load models with the same QP as pre-training parameters instead of the method with the lower QP. For instance, the well-trained baseline model with the first QP is loaded to train the first QP of the proposed perceptual model. Specifically, the fourth column in Fig. \ref{fig:2}. The perceptual quality of images compressed by models trained using the common training strategy is significantly lower than images compressed by models trained using our proposed training strategy.
\subsection{Generalization}\label{sec:3.3}
To verify the generalization of the proposed method, we apply it to another learned image compression scheme, \emph{i.e.}, Cheng \cite{cheng2020learned}, for comparison. For simplicity, we only conduct subjective evaluation and the results are shown in Fig. \ref{fig:4}, our method has more details and better visual quality than Cheng \cite{cheng2020learned}. Moreover, the method is plug-and-play and can be used in most of the existing learned image compression schemes.

% \section{Conclusion}\label{sec:4}
% We start from the fact that the HVS can't perceive the distortion that is under the JND threshold and propose a JND-based perceptual quality loss to replace the MSE loss for optimizing the learned image compression scheme toward high perceptual quality. Different from the distortion caused by the traditional hybrid perceptual image compression which is a constant, the distortion caused by the learned image compression keeps changing during the training process. To better assess perceptual loss during training, we develop a distortion-aware adjustor, which can assess the different amounts of distortion with different levels of JND. Besides, to better fine-tune the baseline with the proposed perceptual loss, we introduce a training strategy. All these designs make the proposed method outperforms the baselines.

\section{Conclusion}\label{sec:4}
We start with the fact that the JND can be used for guiding the distortion assignment for the compressed images, and propose a JND-based perceptual quality loss to replace the MSE loss for optimizing the learned image compression scheme toward high perceptual quality. Different from the distortion caused by the traditional hybrid perceptual image compression which is a constant, the amount of distortion caused by the learned image compression keeps changing during the training process under different QPs. To better assign different amounts of the distortion distribution of the compressed image, we develop a distortion-aware adjustor. Besides, to better fine-tune the baseline with the proposed perceptual loss, we introduce a training strategy. All these designs make the proposed method outperforms the baselines.

\bibliographystyle{IEEEbib}
\bibliography{refs}

\end{document}